# If deep learning is the answer, then what is the question?

Andrew Saxe, Stephanie Nelli and Christopher Summerfield

Department of Experimental Psychology
University of Oxford
Oxford, UK

Correspondence:
andrew.saxe@psy.ox.ac.uk
stephanie.nelli@psy.ox.ac.uk
christopher.summerfield@psy.ox.ac.uk

## Abstract

Neuroscience research is undergoing a minor revolution. Recent advances in machine learning and artificial intelligence (AI) research have opened up new ways of thinking about neural computation. Many researchers are excited by the possibility that deep neural networks may offer theories of perception, cognition and action for biological brains. This perspective has the potential to radically reshape our approach to understanding neural systems, because the computations performed by deep networks are learned from experience, not endowed by the researcher. If so, how can neuroscientists use deep networks to model and understand biological brains? What is the outlook for neuroscientists who seek to characterise computations or neural codes, or who wish to understand perception, attention, memory, and executive functions? In this *Perspective*, our goal is to offer a roadmap for systems neuroscience research in the age of deep learning. We discuss the conceptual and methodological challenges of comparing behaviour, learning dynamics, and neural representation in artificial and biological systems. We highlight new research questions that have emerged for neuroscience as a direct consequence of recent advances in machine learning.

## 1. Introduction

Recent years have seen a dramatic resurgence in optimism about the progress of AI research, driven by advances in deep learning[1]. Deep learning is the name given to a methodological toolkit for building multi-layer (or deep) neural networks that can solve challenging problems in supervised classification[2], generative modelling[3], or reinforcement learning[4,5]. Neuroscience and AI research have a rich shared history[6], and deep networks are now increasingly being considered as promising theories of neural computation. The recent literature is studded with comparisons of behaviour and brain activity in biological and artificial systems[7-21], summarised in a growing number of review articles[22-30]. In this *Perspective*, we assess the opportunities and challenges presented by this new wave of intellectual synergy between neuroscience and AI research.

## 2. Neoconnectionism?

The idea that neural networks can serve as theories of neural computation is not new. During the parallel distributed processing (PDP) movement of the 1980s, psychologists and computer scientists proposed neural networks as solutions to key problems in perception, memory and language[31]. Contemporary deep networks resemble scaled-up connectionist models, and recent advances in machine learning are also heavily indebted to the ubiquity of digital data and the relatively low cost of computation in the 21st century[26]. It might thus be tempting to dismiss current excitement around deep learning models for neuroscience as a rehashing of earlier ideas, owing more to slow churn of scientific fashion than to genuine intellectual progress. However, many researchers believe that deep learning models have the potential to radically reshape neural theory, and to open new avenues for symbiotic research between neuroscience and AI research[23,32-34]. This is because contemporary deep networks are different from their connectionist ancestors in a crucial way: their learning is grounded in quasi-naturalistic sensory signals, such as image pixels[13] or auditory spectrograms[15], rather than their input and output units being hand-labelled by the researcher. Contemporary deep networks can thus learn 'end-to-end' in a sensory ecology that resembles our own: natural sounds and scenes for supervised learning and generative modelling, and 3d environments with realistic physics for deep reinforcement learning. This advent of end-to-end models of biological function has allowed researchers to attempt to model, for the first time, the *de novo* emergence of neural computations that are capable of solving real-world problems.

One major line of research has examined the representations formed by supervised deep networks that are trained to label objects in natural scenes[2] (Fig. 1). A striking observation is that biologically plausible neural representations can emerge in networks that combine gradient descent with a handful of simple computational principles[29]. When deep networks are endowed with local receptivity, convolutions, pooling and normalisation, the early layers acquire simple filters for orientation and spatial frequency[2], just like neurons in area V1; whereas in deeper layers, the distributions and similarity structure of neural representations for objects and categories resemble those in the primate ventral stream[12-14,19]. Representational equivalence may be stronger in more accurate networks[13], and network activations even can be harnessed for novel image synthesis, allowing causal assays of the predictive links between artificial and biological networks[35,36]. One corollary of these findings is that the sophisticated behaviours and structured neural representations observed in humans and other animals might emerge from a limited set of computational principles, as long as the input data are sufficiently rich, and the network is appropriately optimised[25,37].

## 3. Deep learning as a framework for neural computation

This claim has potentially profound implications for neuroscience. It has already prompted calls for systems neuroscientists to refrain from building theories that impose intuitive functional significance on neural circuits by fiat, and instead to study the computations that emerge spontaneously during the training of deep networks[23,32-34]. This so-called *deep learning framework* bypasses the handcrafting in classical neuroscience, where for example constraints are placed on neural encoding by assuming a shape for tuning curves, or population dynamics are explicitly engineered via the wiring diagram of excitatory and inhibitory neurons. Instead, the role of the researchers is now principally to specify the overall architecture, the learning rule and the cost function; control is relinquished over the microstructure of computation, which instead emerges organically over the course of training[33]. An extension of this argument draws an analogy between optimisation over computation in neural networks, and over biological forms by evolution: in both cases, interpretable functional adaptations emerge without meaningful constraints being imposed on the search process[32]. In other words, it is claimed that neural systems are fundamentally uninterpretable, and structured theories of perception and cognition are "just so stories" that reflect more closely the researcher's quest for meaning than the reality of neural computation[32].

One appealing aspect of the deep learning framework is that it relieves researchers of the burden of exhaustively documenting and interpreting the coding properties of single neurons. As methodological advances have permitted simultaneous recordings from large numbers of neurons[38], a doctrine has emerged by which neural representation is dynamically multiplexed across populations[39]. From this perspective, single neurons code for multiple experimental variables and their interactions[40-43], exhibiting the nonlinear, mixed selectivity that is also a hallmark of units in deep networks[44]. This tendency seems to be most pronounced in higher cortical areas, such as the parietal and prefrontal cortex, that support working memory and action selection[40-42]. In these regions, the coding properties of single neurons can be highly heterogenous and vary in mystifying ways over the course of a given trial[40,41,45]. However, when neural activity is examined at the population level, for example by using dimensionality reduction, neural patterns emerge that meaningfully distinguish experimental variables[46,47].

Another key observation is that these patterns of population activity can be recreated when the same analysis is applied to unit activations in recurrent neural networks trained to evaluate time-varying decision evidence[46-48], judge the length of a time interval[49,50], or maintain information over a delay period[51-53]. Accordingly, a major theoretical perspective is emerging that proposes deep recurrent neural networks as computational theories for sensorimotor integration and working memory processes. In the domain of working memory, a particularly interesting new line of research has used recurrent networks to ask when codes for stored information should be static or dynamic, addressing a key question in systems neuroscience[53]. This work has contributed to claims that it is futile to characterise the coding properties of individual cells or infer how they participate in computation[40]. Instead, it is argued, the computational model is only explainable at the aggregate level of the population, which is ultimately driven by the structure of the network and the way it is optimised.

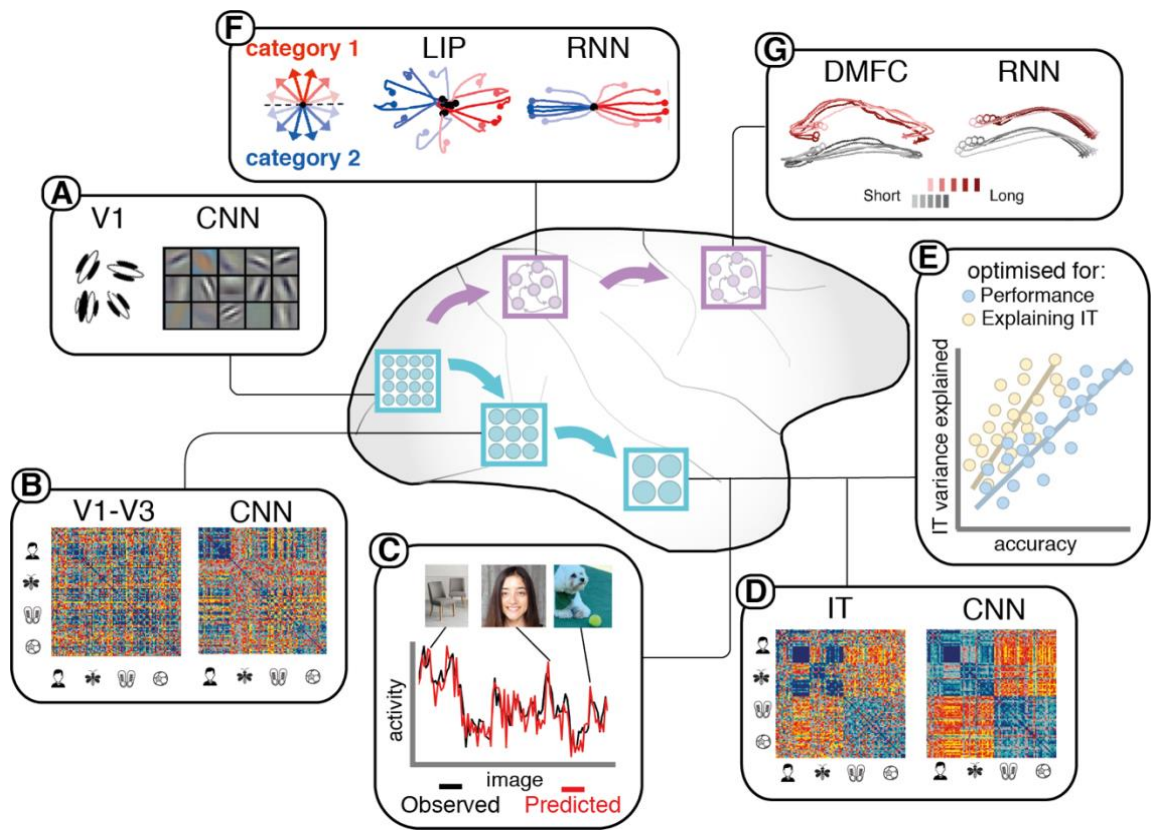

Figure 1. Representational equivalence between neural networks and the primate brain. **A.** Left: schematic illustration of simple and complex cell receptive fields from mammalian V1. Right: example filters learned in the first hidden layer of a deep convolutional neural network. [REF 2] **B.** Example representational similarity matrices illustrating the similarity in population activity evoked by objects in early visual areas of the primate brain (recorded with electrophysiology) and in the intermediate layers in a deep convolutional neural network. [REF 14] **C.** Schematic neural firing rates in response to a series of natural images (black trace; examples above) and activity predicted as a linear transform of neural network activity (red trace). [REF 13] **D.** Representational similarity matrices as in B except comparing inferior temporal (IT) cortex with the final layers of a convolutional neural network. [REF 14]. **E.** Schematic showing relationship between variance explained in IT signals and classification accuracy for pseudo-randomly generated neural networks that are trained either to maximise classification performance (blue dots) or variance explained in neural signals (yellow dots) [REF 13]. **F.** Left: state space analysis of neural signals from macaque area LIP recorded during a dot motion categorisation task. Red and blue lines show different motion directions belonging to opposing categories, trajectories are plotted in two dimensions. Right: same analysis conducted on hidden units of a recurrent neural network [REF 47]. **G.** Left: state space analysis performed on neural signals recorded from macaque dorsomedial prefrontal cortex during performance of fast or slow interval reproduction task, plotted in three dimensions. Right: same analysis conducted on hidden units of a recurrent neural network [REF 50].

## 4. Limitations of the deep learning framework

The deep learning framework proposes powerful new tools for modelling the bewildering volumes of data that are now routinely recorded in systems neuroscience labs. However, we hope that enthusiasm for deep networks as computational models will be tempered with a sober consideration of how they can be usefully deployed to understand neural mechanism and cognitive function. In other words, if deep learning is the answer – then what are the questions that neuroscientists should ultimately be asking?

Unfortunately, as currently articulated, the deep learning framework offers only the scantiest roadmap for neuroscience research[23,32-34]. If neural computation emerges uncontrollably through blind, unconstrained optimisation, then how can neuroscientists formulate new, empirically testable hypotheses about neural mechanism? Such hypotheses are argued to take the form of design choices about learning rules, regularisation principles, or architectural constraints in deep networks[33]. There is some evidence that more judicious design choices for deep networks may permit a closer match to biology[29]. For example, adding recurrent connections improves the fit to neural data[18], especially for more challenging natural images, and at later post-stimulus time points[17], whereas including a biologically plausible front end (a 'retina net') encourages the formation of realistic coding properties, including cell types typically found in the thalamus[54]. In general, however, we lack overall guiding principles for making such design choices. In machine learning research, networks are rarely built with biological plausibility in mind, and so there is relatively little prior guidance in how they might be used to model neural systems. Moreover, understanding the mapping from design to performance in deep networks is challenging, which is presumably why AI has a relatively poor track record in conducting interpretable or overtly hypothesis-driven research, preferring to focus instead on *whether* the system works rather than *why* it works[55].

At worst, the deep learning framework seems to broadside neuroscience with an existential threat. The novel research programme asks researchers to document how different architectures or algorithms can encourage deep networks to form semantically meaningful representations or exhibit complex behaviours, like humans and other animals. This endeavour sounds suspiciously similar to contemporary AI research itself. As such, the project does not seem to build upon the comprehensive understanding of neural computation that has been furnished by decades of research into biological brains. Rather, it seems to propose sweeping this knowledge away, merging the goals of theoretical neuroscience with those of contemporary AI research.

## 5. Are deep networks promising theories of object recognition?

The deep learning framework is built upon the proposal that neural networks learn representations and computations that resemble those in biological brains. However, it is possible that the equivalence between deep networks and animal brains has been overstated[22]. Indeed, comparing the multivariate representations in brains and neural models is fraught with statistical challenges[56]. Currently, one popular approach is to learn a linear mapping from neurons to network units, and to evaluate the predictive validity of the resulting regression model in a held-out dataset. Adopting this approach for image classification, the highest-performing deep networks can explain an impressive 60% of the variance in neuronal responses in primate area IT. However, neural networks that perform substantially worse at image classification explain just 5% less[57]. Indeed, directly comparing predictive accuracy (on BOLD signals) for trained and untrained (initialised to random) networks, the difference is quite small – on the order of 5-10% accuracy difference for most visual regions[19]. It is often forgotten that landmark studies on which claims of representational equivalence are based actually used untrained deep networks[13].

Testing whether neural signals are an affine transform of model activations is a good start, but such a relationship could exist even if the neural patterns differ wildly in terms of sparsity or dimensionality. Stricter tests of shared coding are provided by methods that restrict the freedom of the mapping function, such as representational similarity analysis (RSA)[58]. RSA discloses a superficial resemblance between brains and networks but it is hard to tell whether this agreement is driven principally by stimuli that are physically similar, such as faces, and the differing patterns evoked by animate vs. inanimate objects[14]. An important goal for future research, thus is going to be to more rigorously assess the status of the claim that deep networks and biological brains learn equivalent neural codes.

Another way to test the equivalence between biological and artificial systems is to study their behaviour, i.e. to examine how their response patterns evolve over the course of training, and how they can be experimentally manipulated. This is vital because the computations in a neural system can only be understood in the context of the behaviour they produce[59]. Revealingly, humans and machines make sharply different sorts of errors in assays of object recognition. In one study, networks were prone to confuse object classes that humans and even monkeys could safely tell apart, such as dogs and guitars, and patterns of confusion among individual images were shared by humans and macaques, but not deep networks[11]. Similarly, humans generalise far better than deep networks to images that have been perturbed by adding pixelated noise, or bandpass filtering[9], and are less prone to be fooled by deliberately misleading images[7,21,22]. There is a widespread view that biological vision exhibits a robustness that is currently lacking in supervised deep neural networks.

More generally, animal behaviour is richly structured, in theory permitting researchers to make systematic comparisons with machine performance[60]. For example, animal decisions are subject to stereotyped biases, but also irreducibly noisy[61]; animals are flexible but forgetful, behaving as if memory and control systems were capacity limited[62], and the rate and effectiveness of information acquisition depends strongly on the structure and ordering of the study materials[63]. Mature theories of biological brain function should be able account for these phenomena, and we hope that future deep network models will be held to this standard.

## 6. Theory and understanding of deep learning models

The computations performed by deep networks are enmeshed in millions of trainable parameters. It is not surprising, thus, that they have been dubbed "black boxes". Despite this complexity, in neural networks we can access every synaptic weight and unit activation over the course of learning, a feat that remains impossible in animal models. These considerations raise thorny questions concerning the utility of deep networks as neural models, and more generally, raise the question of what it means to "understand" a neural process via a computational model[34].

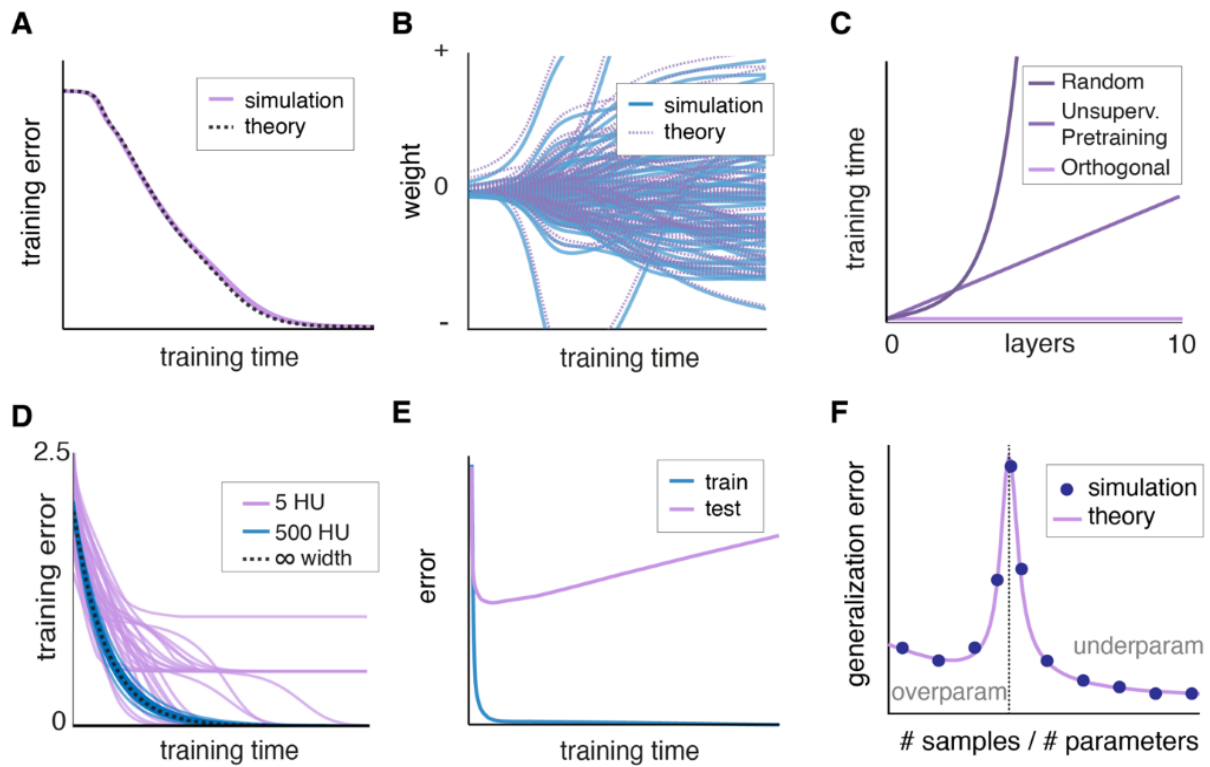

Figure 2. Understanding deep networks using idealized models. **A-B.** By simplifying the neural nonlinearity, deep linear networks permit exact solutions to training error dynamics **(A)** and weight dynamics **(B)** for certain initializations [REF 65]. Full simulations from small random weights track these dynamics closely. **C.** Training speed in deep linear networks depends on initialization. Small random weights scale exponentially in depth, unsupervised pretraining scales linearly, and orthogonal initializations are depth-independent. **D.** Training error of nonlinear networks of different sizes trained on the XOR binary classification task. Networks with few hidden units exhibit complex trajectories, often ending at nonzero error. Large networks reliably find a solution and take similar trajectories. This trajectory can be described analytically in the infinite width limit under a particular initialization regime [REF 73-74]. **E.** Analytical training and testing dynamics for a 'student' network learning from a 'teacher' network, permitting analysis of generalization performance on novel examples and the overtraining phenomenon [REF 67-69]. **F.** Analytical predictions for the generalization error after extensive training in the high-dimensional regime where data is scarce relative to the number of weights. Generalization error peaks at the transition from over- to under-parameterization [REF 69-70].

Thus far, neuroscientists have preferred to employ off-the-shelf deep convolutional networks as neural models[29]. However, collaborations between theoretical neuroscientists, physicists and computer scientists have paved the way for a new approach that uses idealised neural network models to understand the mathematical principles by which they learn[64], and deploys the results to predict or explain phenomena in psychology or neuroscience[65]. For this endeavour to be tractable, deep network models must be simplified, for example by employing linear activation functions ("deep linear" networks)[66], structured environments[67,68], or by studying limit cases, such the limits of infinite width or depth, the high-dimensional limit[69,70], or the shallow limit[64] (Fig. 2). Paradoxically, these infinite-size networks are often more interpretable than those with fewer units, because their learning trajectory is more stable and not prone to be waylaid by bad local minima[68,69,71]. Some network idealisations have offered exact solutions for the learning trajectories that every single synapse will follow[72-74], and answered perplexing questions about network behaviour: for example, why learning often involves transitions between quasi-discrete stages, why deep networks are often slower to train, or why an initial epoch of layer by layer statistical learning ("unsupervised pretraining") can accelerate future learning with gradient descent. This work questions the notion that deep neural networks are "black boxes" and promises interpretable neural network models of biological phenomena (Fig. 2).

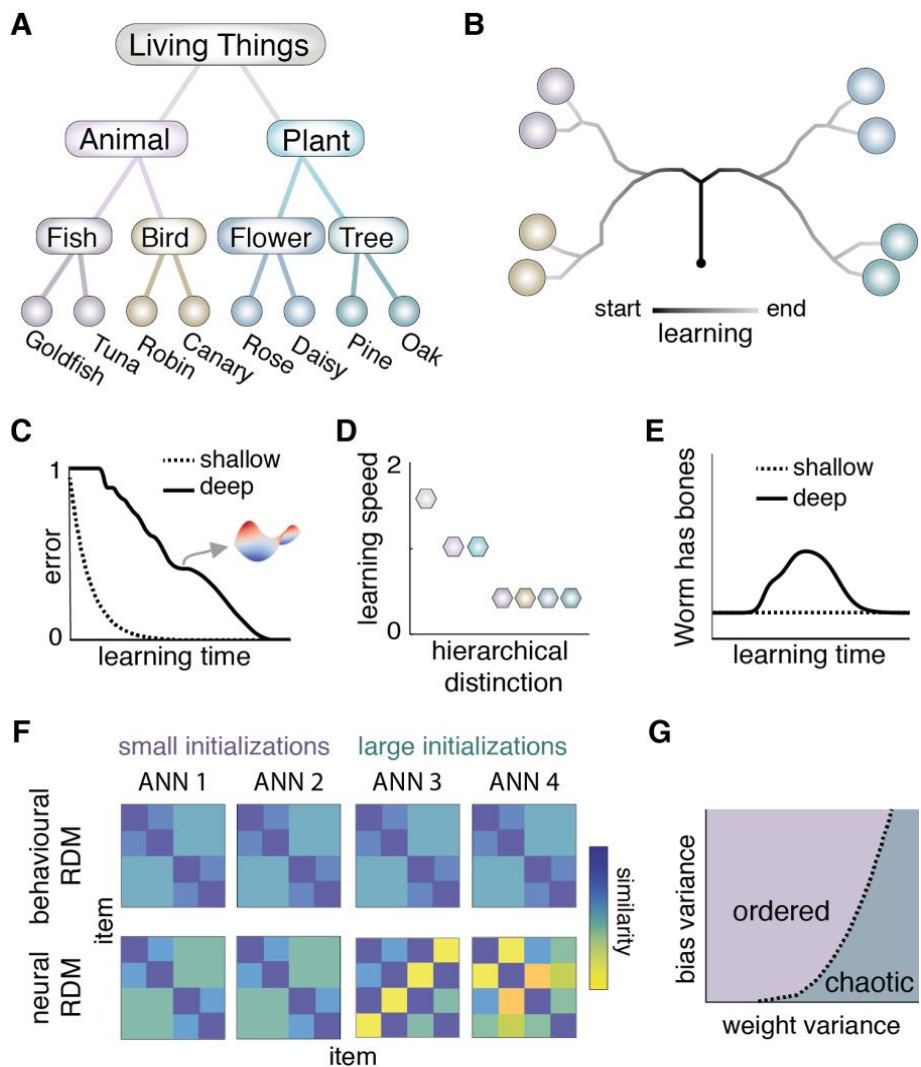

Figure 3. Developmental trajectories in deep linear neural networks. **A.** An idealized hierarchical environment. Items (leaf nodes) possess many properties like "can fly" or "has roots." Nearby items in the tree are more likely to share properties. **B.** Schematic 2D embedding of internal representations for each item over learning in a deep linear network trained to output each item's properties. The network passes through a series of stages in which higher hierarchical distinctions are learned before lower distinctions. **C.** Only deep networks exhibit quasi-stage-like transitions in learning, which arise from saddle points in the error surface. **D.** For a class of hierarchies, learning speed is a decreasing function of hierarchy level, and the network will exhibit progressive differentiation. **E.** Deep but not shallow networks can make transient errors on specific items and properties during learning. **F.** Internal neural representational similarity reliably mirrors behavioural similarity in networks trained from small weights. Networks trained from large weights exhibit correct behaviour but idiosyncratic neural representations. [A-F: REF 65]. **G.** This dependence on initialization can be understood through a transition in learning dynamics between a "feature learning" regime and a "lazy learning" regime [REF 64].

Recently, this approach has been applied to the study of semantic cognition (Fig. 3).[65] During development, children transition through discrete stages in which they rapidly acquire new categories or concepts. Their learning is also highly structured: for example, semantic knowledge is progressively differentiated, as children pick up on broader hierarchical distinctions (animal vs. plant) before finer distinctions (rose vs. daisy) and display stereotyped errors, such as thinking that worms have bones[75]. Deep networks trained on richly structured data are known to exhibit these phenomena[76], but only recently has it been shown that stage-like transitions arise due to saddle points in the error surface, progressive differentiation from the way the singular values of the input-output correlations drive learning over time, and semantic illusions from pressure to sacrifice accuracy on exceptions to meet the global supervised objective[65]. Moreover, it can be shown that these phenomena are an inevitable consequence of depth itself, arising in deep linear networks but not shallow networks even though the two classes of model converge to identical terminal solutions. This highlights the importance for neuroscientists of studying *learning dynamics*, i.e. the trajectory that learning takes, rather than simply examining representations in networks that have converged.

One potential concern is that insights acquired in this way might not scale, because models are idealisations that eschew the messy complexity of state-of-the art deep networks and make assumptions that are false for biology (e.g. linear transduction or layers of infinite width). However, we argue that neural theory is well served by analytic formulations of complex phenomena that give rise to specific, falsifiable predictions for neural circuits and systems. We hope that neuroscientists will incorporate reductions of deep network models into their canonical set of neural theories, rather than blindly seeking correspondences between brains and fully-fledged deep learning systems that offer little hope of being understood.

## 7. Learning rules for sensory systems

Research using idealized neural networks offers the hope that we can understand how learning occurs in biological brains. But specifically, what research questions should we ask? What empirical phenomena might deep networks predict that conventional models from classical neuroscience might not?

Psychologists and neuroscientists have traditionally debated the extent to which perceptual representations are prespecified by evolution or learned via experience[77]. For example, it remains controversial whether primate face representations are innate or acquired[78,79]. The deep learning framework reframes this debate by asking how neural codes emerge from different learning principles. One strong candidate is supervised learning with gradient descent, in which representations are sculpted by feedback about the label, name or category associated with a sensory input[29]. Thus far, supervised models have been the major focus of comparisons between deep networks and biology[29]. However, a long tradition in neuroscience emphasizes unsupervised principles such as Hebbian learning, or argues that representations are formed by a pressure to accurately predict the spatially or temporally local environment under an efficiency constraint[80-82]. Indeed, recent deep generative models show a remarkable ability to disentangle complex, high-dimensional signals into their latent factors under this self-supervised objective[83,84]. Finally, a successful AI model that has yet to impact neuroscience proposes instead that representation formation is driven by the need to accurately predict the motivational value of experience[4].

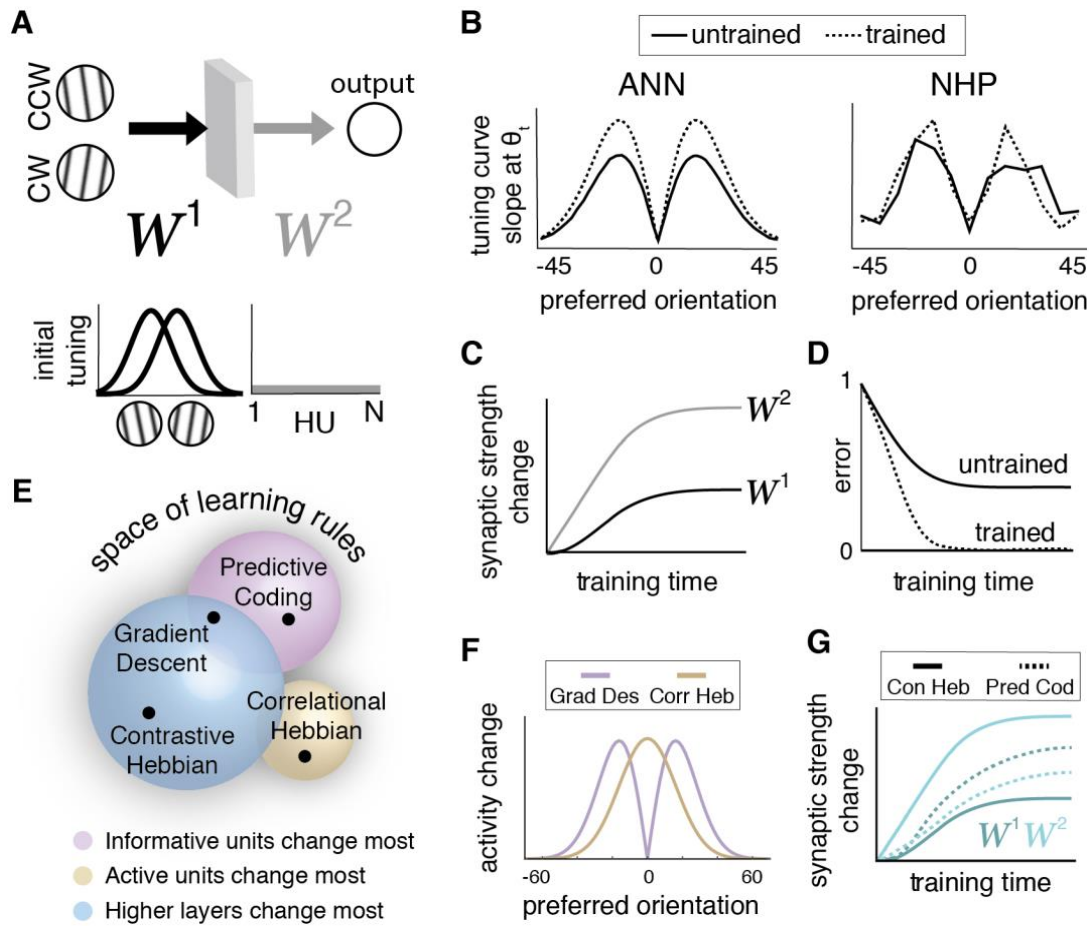

Figure 4. Testing principles of learning using perceptual learning paradigms. **A.** A simple deep network model of perceptual learning. Clockwise or counter clockwise oriented visual inputs flow through two layers of weights to an output layer that reports direction of rotation. Initially, the input layer weights have bell-curve tuning to orientation and the output layer weights are untuned. **B.** Schematic of tuning curve slope changes due to gradient descent learning in the model (left) and primate V1 (right). **C.** Schematic of total synaptic change in input weights and output weights over training. The higher layer changes more. **D.** Schematic of behavioural performance transfer to an untrained retinal location over learning. Early learning transfers well but late learning is retinotopically specific. **E.** A schematized conceptual space of learning rules, in which specific learning rules are points. Experimental observations may be consistent with regions of this space (coloured circles) containing several learning rules. Intersecting many constraints can begin to narrow the set of candidate learning algorithms. **F.** Schematic predictions of gradient descent—for which the most informative neurons change most—and correlation Hebbian learning, for which the most active neurons change most. **G.** Schematic predictions of contrastive Hebbian learning, for which higher layers change more than lower, and a predictive coding scheme for which lower layers change more than higher.

One way to evaluate these schemes is to compare their ability to furnish deep networks with rich representations and complex behaviours when exposed to naturalistic data. To date, most successful candidates use some form of gradient descent. However, standard supervised models, such as those popular for explaining primate object recognition, seem to require improbable quantities of labelled data – unlike human infants, who gain sophisticated object understanding even before language is acquired. Another challenge for networks trained with gradient descent is the problem of assigning credit among parameters. Deep networks cascade simple operations across layers to permit complex input-output transformations, for example learning a hierarchy of detectors for object edges, parts and wholes in successive layers. Whilst this divide-and-conquer strategy maximizes representational power, it demands that a change at one synapse accounts for how this adjustment will propagate through the rest of the network. A grand challenge for neuroscience is to test whether learning in the brain can in fact assign credit across the neural hierarchy, and if so, to identify a biologically realistic implementation, i.e. one where updates are local, and forward and backward connectivity in the network is not required to be symmetric. While credit assignment was once thought to be biologically implausible, we now have a growing set of candidate implementations in need of empirical tests[85,86].

Learning principles also make divergent predictions about how representations should emerge or change during prolonged training. This opens the door to studies of perceptual learning that can attempt to confirm or refute these predictions[87]. For example, Fig. 4 shows the predictions of a neural network model trained to classify tilted gratings with gradient descent, under the assumption that input-layer units have initially bell-shaped tuning curves[66]. Extant neural and behavioural phenomena emerge seamlessly from the model, such as stronger sharpening of the tuning functions of the most informative neurons, earlier and stronger representational changes in higher cortical stages (i.e. deeper layers) during training, greater proneness to transfer of coarse than fine discrimination tasks across space to new retinal locations, and transfer of fine discrimination tasks early but not late during training. These phenomena, which are characterized mathematically in deep linear networks, also occur in the nonlinear case[66,87].

Critically, other learning principles make qualitatively different predictions (Fig. 4E-G). For example, under correlational Hebbian learning, the most active (rather than most informative) neurons change the most. Contrastive Hebbian learning causes the higher layer to change far more than the lower layer, whereas the converse is true for a predictive-coding scheme with top-down negative feedback. Whilst these constraints do not pin down the exact algorithm at work in perceptual learning, they offer collective evidence about the learning principles likely to be at work in biology. More generally, this approach opens the door to a new programme of experiments in which principles of learning are interrogated by measuring the dynamics of representational change across cortical stages using macroscopic imaging techniques such as fMRI or wide-field calcium imaging.

## 8. Deep learning principles for cognition

Deep neural networks excel at classifying complex inputs into distinct classes like objects or words. Equally important, however, is what comes next: we link objects and items into diverse knowledge structures that describe our world. We know, for instance, that a dog can bark and that a maple is a type of tree. Moreover, we form semantic categories from multimodal features, connecting the written and spoken name for an object with its shape, odour, and texture. This conceptual knowledge of the world transcends physical appearance, interlinking diverse and even unobservable object properties (for instance, that a dog has a spleen). The abstractions we acquire over the course of development form the building blocks for flexible generalization and higher-level cognition in maturity[88].

Evaluating deep learning insights beyond the realm of perceptual tasks is a key open opportunity for neuroscientists. The behaviour of humans and other animals is governed by a rich array of cognitive functions, including modular memory processes, attentional and task-level control, and neural systems for navigation, planning, mental simulation, reasoning, and abstract inference. These cognitive functions are implemented in a regionally specialised brain, in which a patchwork of subcortical and allocortical structures interconnects with both granular and infragranular cortical zones, each housing unique cell types and circuits. If we are committed to deploying deep learning models as theories for biology, then we need to take seriously the question of how such elaborate structure in cognition and behaviour emerges from end-to-end optimization. How do humans learn abstract representations, divorced from physical object properties? How do we assemble knowledge into relational structures like trees, rings, and grids? How do we compose new behaviours from existing subcomponents? How do we rapidly acquire and generalise new memories? These are important questions for AI researchers as well, and indeed, some have expressed a hope that machine learning will soon offer more powerful models in which higher cognitive functions emerge naturally via a "blind search" process, allowing neuroscientists to sidestep the problem[32]. Indeed, recent advances in AI research have followed the successful fusion of deep learning with other methods, such as reinforcement learning[4], context-addressable memory[89,90], or Monte-Carlo tree search[5,91], demonstrating a proof of concept for end-to-end learning in complex cognitive architectures. However, we argue that a more fruitful research agenda for neuroscientists builds off the work of past decades, in which researchers have experimentally dissected cognitive systems, in many cases providing a detailed, computationally grounded account of their function. For example, we understand a great deal about the navigation system in the rodent medial temporal lobe[92], the motor system in song birds[93], or the saccadic system in the macaque monkey[94]. We argue for a research programme that embraces the deep learning framework but seeks to address concrete questions about theory and implementation that are recognisable to neuroscientists and cognitive scientists.

## 9. Abstraction and generalisation

Deep networks excel when data is abundant, and training is exhaustive. However, they struggle to extrapolate this knowledge to new environments populated by previously unseen features and objects. Humans, by contrast, seem to generalise effectively[28]. For example, most people can navigate a foreign city where the language, coinage and customs are unfamiliar, because they understand concepts such as *greeting, taxi* and *map*. A popular view is that deep networks fail to transfer because they do not form neural codes that abstract over physically dissimilar domains. Building deep networks that can generalise in this way would be a major milestone for machine learning. This provides an incentive for neuroscientists to study how biological brains encode, compose and generalise abstract knowledge[95-97].

Unfortunately, key methodological challenges arise for neuroscientists seeking to address this question. Firstly, it is unknown whether experimental animals such as rodents and macaques (or even our closer primate cousins[98]) have evolved neural mechanisms that permit the strong, flexible transfer of knowledge that characterise human intelligence. It is thus unclear whether invasive tools for recording and interference (such as electrophysiology or optogenetics) can be used to study generalisation and transfer in animals. To study human abstraction, we are obliged to use macroscopic imaging methods such as fMRI, MEG and EEG, that are less well suited to revealing how computation unfolds in neural circuits. Nevertheless, inventive new ways of using these tools are being developed, allowing researchers to probe replay[99-101], changes in excitatory-inhibitory balance[102,103], or hexagonal (grid) coding[104,105] in human brain signals. Secondly, humans (and other animals) usually enter the laboratory with rich past experiences that sculpt the ways that they learn. This complicates direct comparisons between humans and neural networks, because it is difficult to imbue artificial systems with equivalent priors, or to eliminate human priors using wholly novel stimuli. Thirdly, humans and neural networks learn over very different timescales. For example, deep reinforcement learning systems exceed human performance on Atari video games, but require many times more training than a human player[106].

In an end-to-end learning system, abstract representations need to be grounded in experience. One possibility is that lifelong exposure to huge volumes of sensory data might allow strong invariances to emerge naturally via either supervised or unsupervised learning. In the medial temporal lobe (MTL), which sits at the apex of the primate ventral stream, there is evidence that cells develop physically invariant coding properties. For example, in humans, 'concept' cells code for famous individuals or landmarks, irrespective of whether they are denoted by pictures or words[107]. Echoes of this coding scheme can be seen in other animals, where MTL coding is tied more tightly to allocentric space. For example, hippocampal place cells code for locations in a way that is invariant to the viewpoint and heading direction[108], and in primates, 'schema' cells remap between environments in a way that allows for generalisation over common spatial configurations[109]. These neural codes for high-level concepts can form when different features, objects or locations are repeatedly associated in space or time, for example via Hebbian learning[110]. Indeed, fMRI studies of statistical learning have revealed that neural similarities (e.g. multivoxel pattern overlap) in the MTL recapitulate association strengths for pairs, lines, maps or hierarchies of stimuli[111-118]. Moreover, in an bandit task, the entorhinal cortex is one brain region where a consistent mapping exists between neural patterns and the covariance among stimuli and rewards, irrespective of the physical images involved[119]. Stitching together multiple patterns of association, and learning their structure, could allow animals to learn a comprehensive model of the world that can be used for navigation, inference and planning[120].

In parallel with this growing emphasis on the virtues of model-based computation in neuroscience, machine learning researchers are building powerful deep generative models that are capable of disentangling the world into its latent factors, and recomposing these to construct realistic synthetic images in 3D[84,121,122]. However, to date, wiring these generative models up with control systems to build intelligent agents has proved challenging, despite some promising efforts[90]. Indeed, AI researchers have struggled to build model-based systems that can hold their own against model-free agents in benchmark problems such as Atari[123]. It is paradoxical that this is occurring against a rich backdrop of neuroscience research that emphasises the virtues of model-based inference. In fact, neuroscientists have even begun to unravel how seemingly idiosyncratic coding properties in the MTL and other structures may be hallmarks of a normative scheme for computational efficient planning and inference[124,125]. For example, grid cell codes in the medial entorhinal cortex and elsewhere may be signatures of a neural code that has learned the geometry by which space itself is structured, potentially supporting transfer learning for navigation[125]. There are even hints that this coding scheme may apply to nonspatial as well as spatial domains[105], potentially laying the foundations for a theory of higher-order human reasoning[126]. Although machine learning researchers have noted that lattice-like codes may emerge when deep RL systems are trained to navigate[127,128], they have yet to build on these insights for building stronger AI. More generally, understanding how to simulate biologically plausible model-based computations in a way that is useful to machine learning researchers is a potentially rich intellectual seam that neuroscientists are only just beginning to exploit.

## 10. Neural resource allocation during task learning

Humans and other animals continue to learn across their lifespan. This "continual" learning might allow a human to acquire a second language, a monkey to adopt a new social role, or a rodent to navigate in a novel environment. This is in stark contrast to most current AI systems, that lack the flexibility to acquire new behaviours once they have achieved convergence on an initial task. Building machines that can learn continually, like humans and other animals, is proving one of the thorniest challenges in contemporary machine learning research[129]. Fortunately, however, this question has opened up new avenues for neuroscience research focussed on how biology may have solved continual learning[8,130].

It has long been noted that in neural networks, learning pursuant to an initial task A is often overwritten during subsequent training on task B ("catastrophic interference")[131]. This occurs because a parameterisation that solves task A is not guaranteed to solve any other task, and so during training on task B, gradient descent drives network weights away from the local minimum for task A. It occurs even when the network has sufficient capacity to perform both tasks, because simultaneous (or "interleaved") training allows the discovery of a setting that jointly solves tasks A and B. In humans, new learning can sometimes degrade extant performance, for example when memorising associate pairs A-C after having encoded pairs A-B, but in general interference effects are far less dramatic than for neural networks[132].

One popular model suggests that mammals have evolved to solve continual learning by using complementary learning systems in the hippocampus and neocortex[130,133,134]. Unlike the cortex, hippocampus can rapidly learn sparse (or "pattern-separated") representations of specific experiences, often called "episodic" memories[135], and these memories are replayed offline during periods of rest or sleep[136]. Hippocampal replay provides an opportunity for virtual interleaving of past and present experience, potentially allowing memories to be gradually consolidated into neocortical circuits in a way that circumvents the problem of catastrophic interference. This theory is supported by a wealth of evidence, including the finding that hippocampal damage leads to a gradient of retrograde amnesia[137], and reports of double dissociations between instance-based memory (or "recollection") in the hippocampus and summaries of past experience (or "familiarity") in neocortex[138]. In more recent years artificial replay of past experiences has emerged as a critical factor that allows deep networks to exhibit strong performance in temporally correlated environments[139], including deep reinforcement learning agents for dynamic video games[4]. Pleasingly, this has allowed theorists to draw a link between computational solutions to continual learning in biological and artificial intelligence[140]. Adaptations of the CLS framework allow it to account for seemingly contradictory phenomena, such as the involvement of medial temporal lobe structures in rapid statistical learning[130].

Whilst evidence grows that offline replay may be important for memory consolidation, the problem of continual learning has provoked new questions for neuroscientists. Is biological learning actively partitioned so as to avoid catastrophic interference? Unlike neural networks, animals do not always benefit from interleaved study conditions (imagine learning the violin and the cello at once). For example, humans who have been trained in a blocked fashion to classify naturalistic stimuli (trees) according to two orthogonal boundaries (their "leafiness" vs. "branchiness") perform better on a later interleaved test (compared to those who experienced the same conditions at training and test)[8]. Other evidence from human category learning implies that human knowledge may be actively partitioned by time and context[141,142]. Indeed, promising solutions to continual learning in the machine learning literature rely on the identification of weight subspaces where new learning is least likely to cause retrospective interference, for example by "freezing" synapses that are more likely to participate in extant tasks[143,144]. These tools are more effective when coupled with a gating process that overtly earmarks neural subspaces for new learning, in a way that resembles top-down attention in the primate neocortex[145,146]. Another intriguing possibility is that unsupervised processes facilitate continual learning in biological systems by clustering neural representations according to their context. Hebbian learning might encourage the formation of orthogonal neural codes for different temporally contexts[147], which in turn allows tasks to be learned in different neural subspaces[146]. The curious phenomenon of "representational drift" (where neural codes meander unpredictably over time)[148] might reflect the allocation of information to different neural circuits in distinct contexts, allowing for task knowledge to be partitioned in a way that minimises interference[149].

A more general question is how biological systems have evolved both to minimise negative transfer (interference) and maximise positive transfer (generalisation) among tasks. One fascinating theoretical perspective argues that the capacity limits inherent in biological control processes are a response to this conundrum[150]. Using simulations involving deep networks, the authors show that shared and separate task representations have mixed costs and benefits, with shared codes allowing for generalisation between tasks at this risk of interference between tasks. They suggest that the brain has found a solution by promoting shared neural codes, which in turns allows for strong transfer, but deploying control processes to gate out irrelevant tasks that might provoke interference. They suggest that this answers the question of why, despite a brain that comprises billions of neurons and trillions of connections, humans struggle with multi-tasking problems such typing a line of computer code whilst answering a question[150].

## 11. Conclusions

Deep learning models have much to offer neuroscience. Most exciting is the potential to go beyond handcrafting of function, and to understand how computation emerges from experience. Neuroscientists have recognised this opportunity, but its exploitation has only just begun. In this *Perspective*, we have tried to offer a roadmap for researchers wishing to use deep networks as neural theories. Our principal exhortation for neuroscientists is to use deep networks as predictive models that make falsifiable predictions, and to use model idealisation methods to provide genuine understanding of how and why they might capture biological phenomena. We caution against using increasingly complex models and simulations which outpace our conceptual insight and discourage the blind search for correspondences in neural codes formed by biological and artificial systems. Instead, we hope that neuroscientists will build models that explain human behaviour, learning dynamics and neural coding in rich and fruitful ways, but without losing the interpretability inherent to classical neural models.

## Acknowledgements

This work was supported by generous funding from the European Research Council (ERC Consolidator award to C.S. and Special Grant Agreement 3 of the Human Brain Project) and a Wellcome Trust Sir Henry Dale Fellowship to A.M.S.